%% file: 0-main_acm.tex
\documentclass[sigconf,anonymous=False]{acmart}
\renewcommand\footnotetextcopyrightpermission[1]{}
\usepackage{algorithm}
\usepackage{algpseudocode}
\usepackage{graphicx}
\usepackage{textcomp}
\usepackage{amsmath}
\usepackage{graphicx}
\usepackage{multirow}
\usepackage{multicol}
\usepackage{threeparttable}
\usepackage{enumitem}
\usepackage{subfigure}
\usepackage{placeins}
\usepackage{booktabs}

\AtBeginDocument{%
  \providecommand\BibTeX{{%
    \normalfont B\kern-0.5em{\scshape i\kern-0.25em b}\kern-0.8em\TeX}}}

\setcopyright{acmcopyright}
\copyrightyear{2018}
\acmYear{2018}
\acmDOI{XXXXXXX.XXXXXXX}

\acmPrice{15.00}
\acmISBN{978-1-4503-XXXX-X/18/06}

\usepackage[utf8]{inputenc}
\usepackage{CJKutf8}
\begin{document}
\begin{CJK}{UTF8}{gbsn}
\title{Large Language Models as Evaluators for Recommendation Explanations}



\author{Xiaoyu Zhang}
\email{zhxy0925@gmail.com}
\affiliation{%
  \institution{Tsinghua University}
  \city{Beijing}
  \country{China}
}

\author{Yishan Li}
\email{liyisha19@mails.tsinghua.edu.cn}
\affiliation{%
  \institution{Tsinghua University}
  \city{Beijing}
  \country{China}
}

\author{Jiayin Wang}
\email{jiayinwangthu@gmail.com}
\affiliation{%
  \institution{Tsinghua University}
  \city{Beijing}
  \country{China}
}

\author{Bowen Sun}
\email{sun-bw22@mails.tsinghua.edu.cn}
\affiliation{%
  \institution{Tsinghua University}
  \city{Beijing}
  \country{China}
}

\author{Weizhi Ma}
\email{mawz12@hotmail.com}
\affiliation{%
  \institution{Tsinghua University}
  \city{Beijing}
  \country{China}
}

\author{Peijie Sun}
\email{sun.hfut@gmail.com}
\affiliation{%
    \institution{Tsinghua University}
  \city{Beijing}
  \country{China}
}

\author{Min Zhang}
\email{z-m@tsinghua.edu.cn}
\affiliation{%
  \institution{Tsinghua University}
  \city{Beijing}
  \country{China}
}

\renewcommand{\shortauthors}{Xiaoyu Zhang, et al.}

\begin{abstract}


The explainability of recommender systems has attracted significant attention in academia and industry. Many efforts have been made for explainable recommendations, yet evaluating the quality of the explanations remains a challenging and unresolved issue.
In recent years, leveraging LLMs as evaluators presents a promising avenue in Natural Language Processing tasks (e.g., sentiment classification, information extraction), as they perform strong capabilities in instruction following and common-sense reasoning.
However, evaluating recommendation explanatory texts is different from these NLG tasks, as its criteria are related to human perceptions and are usually subjective.

In this paper, we investigate whether LLMs can serve as evaluators of recommendation explanations.
To answer the question, we utilize real user feedback on explanations given from previous work and additionally collect third-party annotations and LLM evaluations. 
We design and apply a 3-level meta-evaluation strategy to measure the correlation between evaluator labels and the ground truth provided by users.
Our experiments reveal that LLMs, such as GPT4, can provide comparable evaluations with appropriate prompts and settings.
We also provide further insights into 
combining human labels with the LLM evaluation process and utilizing ensembles of multiple heterogeneous LLM evaluators to enhance the accuracy and stability of evaluations.
Our study verifies that utilizing LLMs as evaluators can be an accurate, reproducible and cost-effective solution for evaluating recommendation explanation texts. Our code is available here\footnote{https://github.com/Xiaoyu-SZ/LLMasEvaluator}.

\end{abstract}

\maketitle

\input{1-introduction.tex}

\input{2-setup.tex}

\input{3-RQ1.tex}

\input{4-RQ2.tex}
\input{5-RQ3.tex}
\input{6-RelatedWork.tex}

\input{7-Conclusion}

\appendix
\bibliographystyle{ACM-Reference-Format}
\bibliography{sample-base}

\end{CJK}
\end{document}

%% file: 1-introduction.tex
\section{Introduction}
Explainability has always been a topic of great concern within the field of recommendation~\cite{ExplainSurveyYongfengZhang,chen2022measuring,sun2020dual,sun2021unsupervised}. 
Researchers explore various methods to help users understand why recommendation systems give certain results. 
Among these methods, the text-based explanation emerges as a kind of prominent and widely-used approach~\cite{DBLP:GenerateLanguage,li2021personalizedTransformer,li2022personalized}. 
Through explanation text, systems can present the reason for recommendations to users in natural language, thereby increasing user trust and comprehensibility of the given results.

An effective evaluation should ensure the explanations truly resonate with users and meet their expectations.
However, with advanced approaches developed for generating explanatory text, assessing their quality remains an issue that has yet to be adequately resolved. 
Existing methods for evaluation can be categorized into three main types: self-report, third-party annotations, and reference-based metrics. 
Conducting user studies to obtain self-reported feedback can most accurately reflect user experience. This approach requires evaluations to be recorded alongside the recommendations and is, therefore, difficult to obtain and use in public datasets. 
Third-party annotations can reflect human feedback and are relatively accessible. 
Still, manual labeling is expensive, time-consuming, and lack of scalability.
Reference-based metrics assess the quality of the target text based on the reference text and offer a standard assessment that is relatively easy to acquire. Common metrics such as BLEU~\cite{DBLP:bleu2022ACL}, and ROUGE~\cite{Rouge2004Lin} calculate the similarity between the generated text and the reference text. 
However, there may not exist reference texts for the scenario (many use reviews as a substitute).
In addition, these textual similarities metrics may not ideally reflect user perception of recommendation explanation~\cite{ExpScore2022WWW,stopbleu}.
These limitations highlight the need to develop evaluation methodologies that are in-line with human experience, easy to acquire, and reproducible.

Recently, the development of large language models (LLM) sheds new light on the evaluation of various neural language generation (NLG) tasks~\cite{kocmi2023LLMEvaluateTransaltion,wang2023automatedpersonaliztion,wang2023chatgpt,fu2023gptscore}. 
Considering that LLMs can follow human instructions and their language modeling ability, LLMs can make adequate evaluations under reference-free settings with appropriate prompt~\cite{GEval2023liu,he2024annollm}.
LLM offers an appealing solution for evaluating the quality of recommendation explanations since it is efficient (lower cost than manual labeling) and widely applicable (almost no dataset limitation).
In a study on the design of explainable recommendation method~\cite{lei2023recexplainer}, researchers introduce LLMs to evaluate the quality of explanations generated.
However, we state that successes in the evaluation of general NLG tasks can not be migrated to the evaluation of explainable recommendations without verification.
Unlike those NLG tasks that are already been examined, evaluating recommendation explanation text is more sophisticated.
The reason is that measuring the quality of recommendations involves a group of diverse goals, e.g. persuasiveness, transparency~\cite{ConflictGoalsOfExplanations}, etc. 
Additionally, a large portion of these goals are related
to the subjective perception of users.
All these factors add to the difficulty of assessing the quality of recommendation explanation texts.
These considerations underscore the need to further explore the feasibility of leveraging LLM as a potential solution for text-based explainable recommendation evaluation.

In this paper, we are concerned with the research question: Can LLM serve as an evaluator of recommendation explanation text?
In particular, we delve into three detailed inquiries:


\textbf{RQ1} Can LLMs evaluate different aspects of user perceptions about the recommendation explanation texts in a zero-shot setting?

\textbf{RQ2} Can LLMs collaborate with human labels to enhance the effectiveness of evaluation?

\textbf{RQ3}  Can LLMs collaborate with each other to enhance the effectiveness of evaluation?

To answer these research questions, we use data from a user study in previous work~\cite{UserPerceptionTois}, including the users' ratings on 4 aspects of the provided explanatory texts, as well as the self-explanation text of the users for the recommended movies.
Our study is based on the premise that the user receiving the recommendation is the ground-truth evaluator for the quality of the explanation.
To study the reliability of LLM annotations, we additionally collect third-party annotations and LLM annotations under human instructions. 
To comprehensively compare between evaluators, we design and apply a 3-level meta-evaluation strategy to measure the correlation between evaluator annotations and user labels on different aspects.
We compare the evaluation accuracy of LLM evaluators with third-party annotations and commonly used reference-based metrics, i.e., BLEU~\cite{DBLP:bleu2022ACL} and ROUGE~\cite{Rouge2004Lin}.

Our main findings are: 
(1) Certain zero-shot LLMs, such as GPT4, can attain evaluation accuracy comparable to or better than traditional methods, with performance varying across different aspects.
(2) The effectiveness of one-shot learning depends on backbone LLMs.
Particularly, personalized cases can assist GPT4 in learning user scoring bias.
(3) Ensembling the scoring of multiple heterogeneous LLMs can improve the accuracy and stability of evaluation.


In summary, our contributions include investigating the feasibility of using zero-LLM as the evaluator for explainable recommendations, discussing possible collaboration paradigms between LLM and human labels, and exploring the aggregations of multiple LLM evaluators.
We propose that LLM can be a reproducible and cost-effective solution for evaluating recommendation explanation text with appropriate prompts and settings. 
Compared with traditional methods, LLM-based evaluators can be applied to new datasets with few limitations.
By introducing this evaluation approach, we aspire to contribute to the advancement of the area of explainable recommendation.

%% file: 2-setup.tex
\section{Experimental Setup}

\label{sec:related_work}

\begin{figure*}[t!]
  \vspace{-5mm}
    \centering
    \includegraphics[width=1\linewidth]{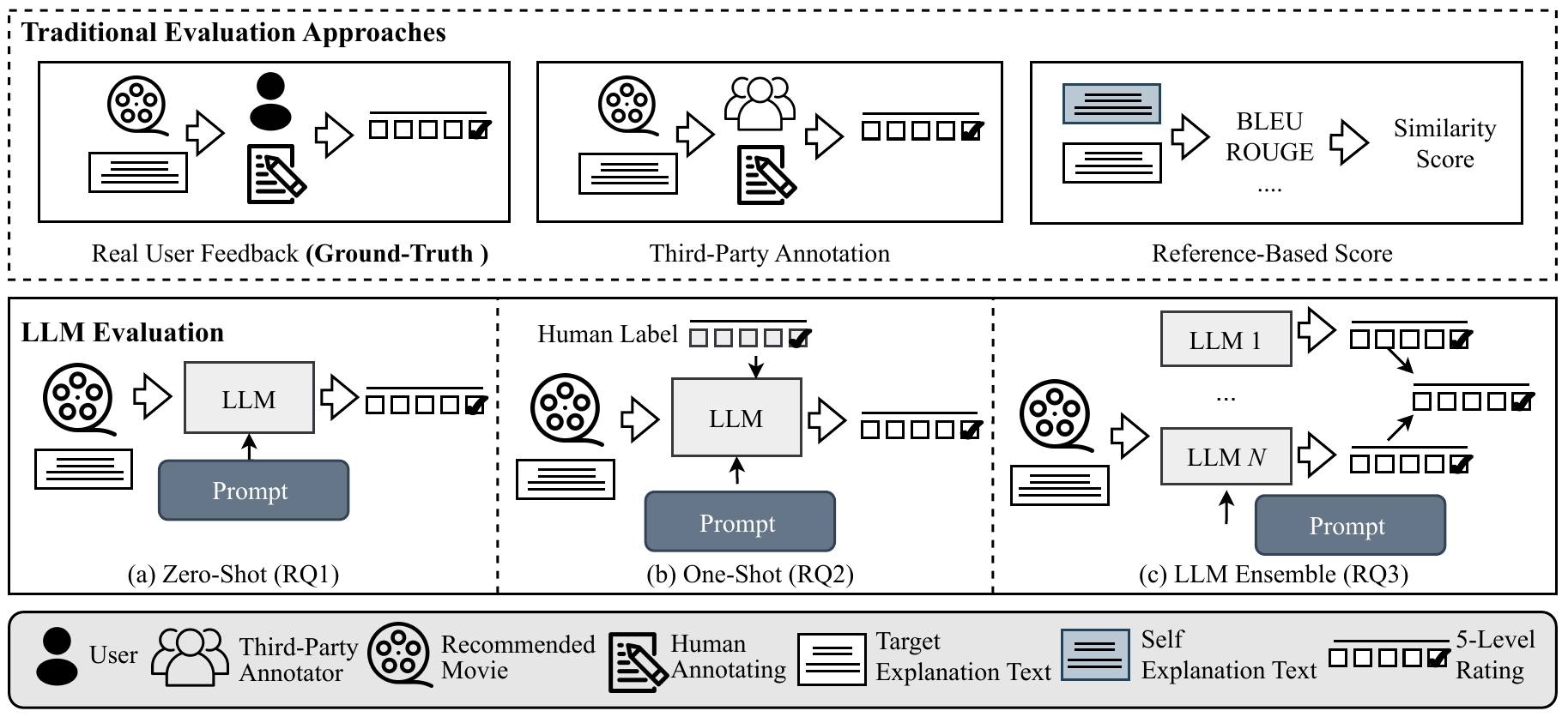}
    \caption{Traditional evaluation approaches vs. utilizing LLMs for evaluations.}
    \label{fig:main}
\end{figure*}

\subsection{Problem Formulation}

We use $\mathcal{U} = (u_1, u_2, ...u_{|\mathcal{U}|})$ to denote the set of users in an RS (Recommendation System).
The RS recommend multiple items to each user $u$, which are defined as $\mathcal{I}_{u} = (i_1, i_2, ..., i_{|\mathcal{I}_{u}|})$. 
When the item $ i \in \mathcal{I}_{u}$ is recommended to the user $u$, explanation texts are generated by a group of generation methods, which are denoted as $\mathcal{G} = (g_1(\cdot), g_2(\cdot), ..., g_{|\mathcal{G}|}(\cdot))$.  $E_{u,i} = g(u,i)$ denotes the explanation text given by $g$ when recommending item $i$ to user $u$.

$f(\cdot)$ denotes the evaluation methods for $E$. 
We assume that users in the system are the most accurate evaluators for explanations of items recommended to them. Their evaluations are denoted as $\mathbf{s_{(u,i,E)}} = f_{u}(i,E)$.
When utilizing $f_{LLM}$ to approximate $f_{u}$, the evaluation given by LLM can be represented as:
$$\hat{\mathbf{s}}_{LLM(u,i,E)} = f_{LLM}(u,i,E) = LLM(u,i,g(\cdot),\mathcal{P})$$
where $\mathcal{P}$ denotes the prompt that contains human instructions. 

Meta-evaluation method $h(\cdot)$ based on correlation metrics are introduced to measure the similarity between $\mathbf{s}$ and $\hat{\mathbf{s}}$. The accuracy of the evaluation given by LLM can be expressed as $h(\mathbf{s},\hat{\mathbf{s}}_{LLM})$.
In our paper, we examine the feasibility of $f_{LLM}$ by analysing the value of $h(\mathbf{s},\hat{\mathbf{s}}_{LLM})$. 
$h(\mathbf{s},\hat{\mathbf{s}}_{t})$ and $h(\mathbf{s},\hat{\mathbf{s}}_{r})$ are used as referent standards, where $\hat{\mathbf{s}}_{t}$ denotes evaluation given by third-party annotators, and $\hat{\mathbf{s}}_{r}$ denotes evaluation given by referenece-based metrics.

\subsection{Data Construction}


\subsubsection{Data overview}
\label{sec:data_overview}
\citet{UserPerceptionTois} create a movie recommendation platform.
It first captures user preferences, and then gives personalized recommendations along with text-formed explanations generated by a series of systems. We utilize two parts of the data: 1) self-explanations written by users, 2) 1-5 ratings of users to explanations generated by different methods in terms of 4 aspects, which are persuasiveness, transparency, accuracy, and satisfaction. 

Formally, 39 participants are included in $\mathcal{U}$ and $\mathcal{I}$ are movies from Movielens Latest dataset\footnote{https://grouplens.org/datasets/movielens/latest/}.
$\mathcal{I}_u$ denotes the itemset recommended to the user $u$, which includes the top-8 movies calculated by BiasedMF~\cite{RendleMF2012}.
$\mathcal{G}$ includes a series of systems used to generate the explanatory texts.
Most of these methods are template-based methods, e.g. user-based method, which generates explanations in the form of: ``[N\%] of users who share similar watching tastes with you like [MOVIE TITLE] after watching it.''
$\mathcal{G}$ also includes the system that directly generates the complete natural language sentence, i.e., peer-explanation written by others. The data includes the self-explanation for each user-item pair.
Since they are written by users themselves, there is no evaluation rating attached.

In summary, the data comprises entries from 39 users, each of whom received recommendations for 8 movies. For each user-movie pair, approximately 8 explanations are generated, resulting in a total of around 2,500 text entries. The data are collected in Chinese.
We take this data as ground truth $\mathbf{s}$ in the experiments.
We additionally collect evaluation scores $\mathbf{\hat{s}}$ given by third-party annotators, LLMs and quantitative metrics.
The evaluation approaches from which the data is derived are illustrated in Figure~\ref{fig:main}.

\subsubsection{Evaluations from users}

For each explanation text $E$ generated, the user is asked to give a 5-scale Likert score to the explanations. The user feedback of the explanation is given on 4 aspects:

\textbf{Persuasiveness}: This explanation is convincing to me;

\textbf{Transparency}: Based on this explanation, I understand why this movie is recommended;

\textbf{Accuracy}: This explanation is consistent with my interests;

\textbf{Satisfaction}: I am satisfied with this explanation.

These 5-scale Likert questions are also utilized in collecting third-party evaluations and LLM evaluations to ensure consistency in the definitions of aspects. 
The user annotation for each explanatory text $\mathbf{s}_{(u,i,E)}$ is a one-dimensional vector of length 4, with elements being integers between 1 to 5.

\subsubsection{Evaluation from third-party annotators}

We employ two annotators to evaluate the explanatory text from the above-mentioned four aspects. The annotators are informed that the user is utilizing a movie recommendation platform and receiving film recommendations along with the reasons for those recommendations. The annotators are asked to score the explanatory text in an item-wise manner.
The third-party annotation for each explanatory text $\mathbf{\hat{s}}_{t(u,i,E)}$ is a one-dimensional vector of length 4, with elements being integers between 1 to 5.

\subsubsection{Evaluation from Quantitative metrics}

We utilize BLEU~\cite{DBLP:bleu2022ACL} and ROUGE~\cite{Rouge2004Lin} to evaluate the explanatory text, taking the self-explanations written by users as reference texts. 
The evaluation score $\mathbf{\hat{s}}_r$ is a float that measures text similarity.

\subsection{LLM as Evaluator}
\label{sec:llmevaluator} 

\subsubsection{LLM Evaluator Construction}

We utilize pre-trained LLMs to provide annotations for each explanation text $E$. The large language model receives the movie name and the corresponding explanation text, accompanied by a prompt $\mathcal{P}$ to describe the context. 
Although $E$ is in Chinese as introduced in Section \ref{sec:data_overview}, we apply $\mathcal{P}$ in English to ensure a standardized prompt design.
The LLM evaluation for each explanatory text $\mathbf{\hat{s}}_{LLM(u,i,E)}$ is a one-dimensional vector of length 4, with elements being integers between 1 to 5.
Our experiments are based on item-wise evaluation, where only one text entry is evaluated at a time. This approach aligns with the setup in the user study.

To investigate the aforementioned RQs, three basic methods are used to construct the LLM evaluator, as illustrated in Figure \ref{fig:main}. 
First, we directly use zero-shot LLMs for evaluation. Then, we consider providing human labels as contextual information to enhance their abilities to learn user subjective perceptions. 
Finally, inspired by the traditional approach of collecting multiple annotators labels and then averaging the results, we similarly ensemble the results of multiple annotations.
When constructing LLM evaluators, we aim to maintain the simplicity and transferability of the approaches. Therefore, we have not leveraged other common methods that require model training, such as fine-tuning.

\begin{figure*}[htbp]
    \centering
    \includegraphics[width=0.9\linewidth]{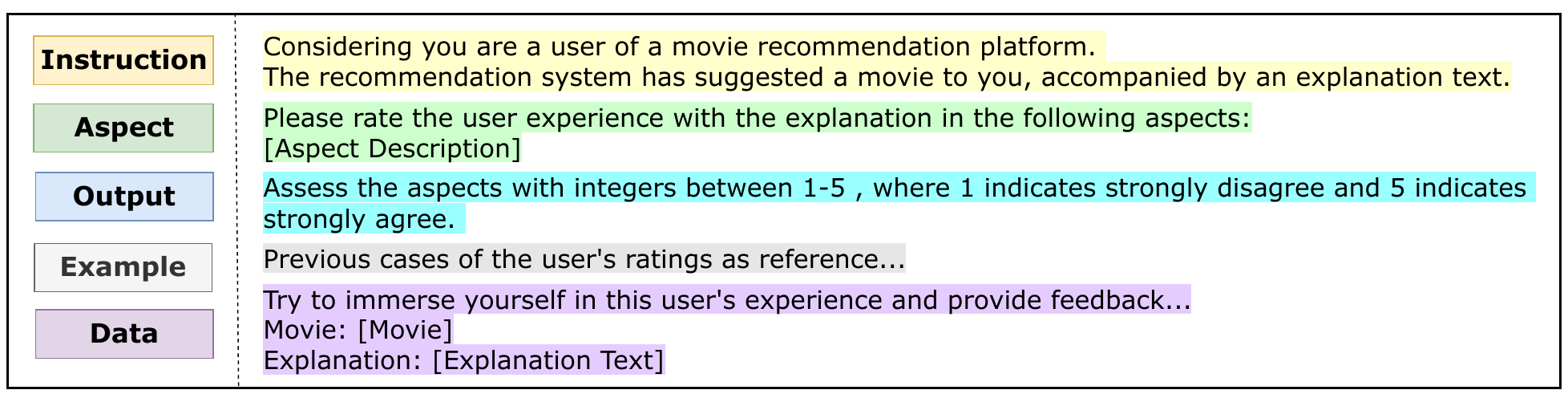}
    \vspace{-2mm}
    \caption{The outline of evaluation prompt templates applied in our study. }
    \label{fig:prompt}
\vspace{-3mm}
\end{figure*}

\subsubsection{Prompt Construction}
\label{sec:prompt}

In the section, we mainly introduce how we construct the prompt $\mathcal{P}$.
The prompt is designed to guide the LLM in evaluating the quality of explanatory text from specified aspects.
It includes several key components, briefed in Figure \ref{fig:prompt}.
We attempted to integrate user information in the prompt but found that this approach did not improve performance and sometimes even decreased it.
Following, we describe in detail the different settings in prompt constructions.

\textbf{Single-Aspect vs. Multiple-Aspect.}
\label{sec:aspect}
The difference between single and multiple-aspect prompts lies in the Aspect part. A multiple-aspect prompt instructs the model to evaluate all four aspects concurrently, while a single-aspect prompt focuses on one aspect at a time. 
Utilizing single-aspect prompts necessitates more times of interactions with LLM compared to multiple-aspect prompts. Experimental results under the two settings are presented and compared in RQ1, Section \ref{sec:RQ1}.

\begin{algorithm}
\caption{Comparison of personalized and non-personalized one-shot learning process}
\begin{algorithmic}[1]
\Procedure{PersonalizedOneShot}{}
    \For{each user $u$ in $\mathcal{U}$}
        \State Randomly choose item $i_0$ from $\mathcal{I}_u$
        \State Collect user evaluations on explanatory text $E=g(u,i_0)$ as $\mathbf{s_0}_g$ for each $g$ in $\mathcal{G}$
        
        \For{each movie $i$ in $\mathcal{I}_u$, each $g$ in $\mathcal{G}$}
            \State $Prompt \gets$ \textproc{ConstructPrompt}($i$, $g(u,i)$, $\mathbf{s_0}_g$)
            \State $Result \gets$ LLM($Prompt$)
            \State Parse $Result$ and process the answer
        \EndFor
    \EndFor
\EndProcedure
\Procedure{Non-PersonalizedOneShot}{}
\State Randomly choose user-item pair $(u_0,i_0)$ 
\State Collect user evaluations on explanatory text $E=g(u_0,i_0)$ as $\mathbf{s_0}_g$ for each $g$ in $\mathcal{G}$
 \For{each user $u$ in $\mathcal{U}$,each movie $i$ in $\mathcal{I}_u$, each $g$ in $\mathcal{G}$}
    \State $Prompt \gets$ \textproc{ConstructPrompt}($i$, $g(u,i)$, $\mathbf{s_0}_g$)
    \State $Result \gets$ LLM($Prompt$)
    \State Parse $Result$ and process the answer
    \EndFor
\EndProcedure
\end{algorithmic}
\end{algorithm}
\textbf{Zero-Shot vs. One-Shot.}
Zero-shot learning presents the task without any example and relies solely on the pre-trained knowledge and reasoning ability of the LLM.
While this may result in a lack of context-specific guidance, the advantage is that it can be applied directly to datasets without manual labeling.

To explore whether humans and LLM can collaborate on evaluation,  we investigate the impact of utilizing human labels as contextual information in LLM evaluation.
Our primary focus is on one-shot learning and includes one human label in the prompt.
To enhance the LLM's ability to learn personalized preferences, we employ personalized one-shot learning. 
This approach involves providing the scoring example from the same user for the target data. Formally, for the user $u$, item $i$ and explanation $E=g(u,i)$, the personalized example is the scoring given to $E'=g(u,i_0)$, which is the explanation generated by the same system $g$ and the same user $u$ on a randomly chosen movie item $i_0$.
The methods help the prompt $\mathcal{P}$ incorporate personal information.
However, a limitation is that it still requires collecting real user feedback, which is expansive and sometimes impractical.
Hence, we also investigate non-personalized one-shot prompts.
In this setting, all examples are from randomly selected $(u_0, i_0)$ pair.
That is, the example can come from another user and are easier to collect in practice.
Formally, for the user $u$, item $i$ and explanation $E=g(u,i)$, the non-personalized example is the scoring given to $E'=g(u_0,i_0)$.
We summarize the personalized and non-personalized one-shot learning procedures in Algorithm 1 to better illustrate the example selection process and their differences.
Experimental results under different example settings are presented and compared in RQ2, Section \ref{sec:RQ2}.

\subsection{Three-Level Meta Evaluation}
\label{sec:metaevaluation}

Good evaluation requires that evaluation efforts
themselves be evaluated~\cite{stufflebeam1974metaevaluation}. 
Employing the suitable meta-evaluation method $h(\cdot)$ is crucial for thoroughly examining the evaluation procedure.
Correlation coefficients, such as Pearson($r$), Spearman($\rho$), and Kendall($\tau$) coefficients, have served as a widely used metric for gauging the similarity in trends between two arrays of ratings or scores~\cite{liu2023calibrating}.
In NLG, the strategy of meta-evaluation can be divided into two levels: Dataset-Level and Sample-Level~\cite{liu2023calibrating}. 
Formally, given a dataset $\mathcal{D}$ consisting of a set of source data $\mathcal{X} = (x_1, x_2, ...., x_{|\mathcal{X}|})$ and generation methods $\mathcal{G} = (g_1(\cdot), g_2(\cdot), ..., g_{|\mathcal{G}|}(\cdot))$ and correlation metric $r(\cdot)$ dataset-level and sample-level meta evaluation are expressed as:

\textbf{Dataset Level } 
\begin{equation*}
    h_D(\mathbf{s},\hat{\mathbf{s}}) = r\left(
    \left(\mathbf{s}_1, \mathbf{s}_2, ... \mathbf{s}_{|\mathcal{X}||\mathcal{G}|} \right),
    \left(\hat{\mathbf{s}}_1, \hat{\mathbf{s}}_2, ... \hat{\mathbf{s}}_{|\mathcal{X}||\mathcal{G}|} \right)
    \right) \\
\end{equation*}

\textbf{Sample Level} 
\begin{equation*}
    h_S(\mathbf{s},\hat{\mathbf{s}}) = 
    \frac{1}{|\mathcal{X}|}\sum_{i=1}^{|\mathcal{X}|}
    r\left(
    \left(\mathbf{s}_{(i,1)}, ... \mathbf{s}_{(i,|\mathcal{G}|)} \right),
    \left(\hat{\mathbf{s}}_{(i,1)}, ... \hat{\mathbf{s}}_{(i,|\mathcal{G}|)} \right)
    \right)
\end{equation*}

Previous studies have not discussed a multiple-level meta-evaluation when assessing the quality of evaluation metrics for explainable recommendations. 
However, considering multiple-level meta-evaluation in explainable recommendations is worthwhile.
This is because the distribution of ground-truth labels or evaluations generated by models may differ between users or user-item pairs.
Consequently, while an evaluation metric might effectively capture trends within specific groups (such as comparing the qualities of a group of texts derived from the same user-movie pair), it may struggle to accurately depict trends across groups (such as capturing certain users' inclination to provide higher ratings).

Therefore, we propose a 3-level strategy for meta-evaluation on recommendation explanations. 
Our motivation is to divide the data into groups at various grain scales and measure the correlation between evaluation results with ground-truth labels within each group.
The three proposed levels are Dataset-Level, User-Level, and Pair-Level.
Together, they provide a comprehensive view for $h$:

\textbf{Dataset Level} correlation calculates the correlation of all scores. The total number of data can be expressed as $|\mathcal{D}| = |\mathcal{U}|\cdot|\mathcal{I}_{u}|\cdot|\mathcal{G}|$.

\begin{equation*}
    h_D(\mathbf{s},\hat{\mathbf{s}}) = r\left(
    \left(\mathbf{s}_1, \mathbf{s}_2, ... \mathbf{s}_{|\mathcal{D}|} \right),
    \left(\hat{\mathbf{s}}_1, \hat{\mathbf{s}}_2, ... \hat{\mathbf{s}}_{|\mathcal{D}|} \right)
    \right)
\end{equation*}

\textbf{User Level} correlation calculates the correlation within the data of each user and then averages it. The total number of data derived from the same user $u$ can be expressed as $|\mathcal{D}_u| = \cdot|\mathcal{I}_{u}|\cdot|\mathcal{G}|$.

\begin{equation*}
    h_U(\mathbf{s},\hat{\mathbf{s}}) = 
    \frac{1}{|\mathcal{U}|}
    \sum_u r\left(
    \left(\mathbf{s}_1, \mathbf{s}_2, ... \mathbf{s}_{|\mathcal{D}_u|} \right),
    \left(\hat{\mathbf{s}}_1, \hat{\mathbf{s}}_2, ... \hat{\mathbf{s}}_{|\mathcal{D}_u|} \right)
    \right)
\end{equation*}

\textbf{Pair Level} correlation calculates the correlation within the data of each user-item pair and then averages it. $|\mathcal{G}|$ explanation texts are derived from the same user-item pair by $g_1(\cdot),g_2(\cdot),...,g_{\mathcal{G}}(\cdot)$.

\begin{equation*}
    h_P(\mathbf{s},\hat{\mathbf{s}}) = 
    \frac{1}{|\mathcal{U}|\cdot|\mathcal{I}_{u}|}
    \sum_{u,i\in \mathcal{I}_u} r\left(
    \left(\mathbf{s}_1, \mathbf{s}_2, ... \mathbf{s}_{|\mathcal{G}|} \right),
    \left(\hat{\mathbf{s}}_1, \hat{\mathbf{s}}_2, ... \hat{\mathbf{s}}_{|\mathcal{G}|} \right)
    \right)
\end{equation*}

Referring to correlations on which level depends on the settings of the task. 
For instance, if an evaluation metric is used to compare between $\mathcal{G}$, it is more important to look at the Pair-Level correlation between the results from the evaluation metric and ground-truth labels. 
When the study requires a measure of satisfaction with the recommendation explanation text by different users of the system, the results of Dataset-Level correlation should be referred to.

We use a concrete example to illustrate the possible variation of the same evaluation metric at different levels.
For instance, our experimental results in Section \ref{sec:RQ1} found that BLEU-4 scoring results correlate poorly or even negatively with ground-truth labels at the Dataset-Level, 
whereas correlations improve at User-Level and Pair-Level. 
This discrepancy arises because BLEU-4 computes the token similarity between the target text and the reference text. 
Some reference texts may contain more commonly used token combinations than others, potentially inflating the BLEU-4 score. 
This introduces a bias into the evaluation process that is unrelated to user perceptions, thereby impacting the Dataset-Level correlation. Full experimental results and analyses can be found in Section \ref{sec:RQ1}.

\label{sec:RQ1}
 \begin{table*}[t!]
  \caption{The 3-level Pearson correlation between the results given by the evaluator and the ground-truth label given by users. 
  Bold fonts denote the best results among all tested evaluators and the
  underlines show the second-best results.}
  \label{tab:main}
  \centering
  \begin{tabular}{p{2cm}|c c c c c}
    \toprule
    \multicolumn{6}{c}{Dataset-Level / User-Level / Pair-Level (\%)}\\
    \midrule
    \multicolumn{1}{c}{Method}& 
    \multicolumn{1}{c}{\textbf{Persuasiveness}}&
    \multicolumn{1}{c}{\textbf{Transparency}} &
    \multicolumn{1}{c}{\textbf{Accuracy}} &
    \multicolumn{1}{c}{\textbf{Satisfaction}} &
    \multicolumn{1}{c}{\textbf{Average}}
    \\



    

    \cmidrule(lr){1-1}
    \cmidrule(lr){2-6}
    Random & -0.55 / 0.52 / 1.81&0.65 / -0.43 / -2.58&-0.41 / 4.12 / 3.98&0.36 / -2.26 / 5.88&0.01 / 0.49 / 2.27 \\

    \midrule
    \multicolumn{6}{c}{{\textit{Reference-based Metric}}}\\
    \midrule

    BLEU-1 & 11.68 / 15.84 / 17.07&10.06 / 12.69 / 14.44&6.43 / 10.71 / 12.18&11.36 / 12.91 / 15.79&9.88 / 13.04 / 14.87  \\
    BLEU-4 & -1.17 / 7.68 / 13.53&-3.47 / 4.13 / 10.24&-4.63 / 4.8 / 8.96&0.61 / 6.86 / 12.09&-2.16 / 5.86 / 11.21 \\
    ROUGE-1-F & 14.16 / 16.39 / 17.56&11.93 / 12.74 / 14.45&8.61 / 11.02 / 12.83&12.87 / 13.2 / 16.23&11.89 / 13.34 / 15.27 \\
    ROUGE-L-F & 14.16 / 16.39 / 17.56&11.93 / 12.74 / 14.45&8.61 / 11.02 / 12.83&12.87 / 13.2 / 16.23&11.89 / 13.34 / 15.27 \\
    \midrule
    \multicolumn{6}{c}{{\textit{Annnotation}}}\\
    \midrule
    Annotator-1 & 19.88 / 18.31 / 16.72&15.66 / 16.18 / 11.31&10.16 / 9.78 / 9.77&14.93 / 13.28 / 12.69&15.16 / 14.39 / 12.62 \\
    Annotator-2 & 21.4 / 21.17 / 20.9& \textbf{25.97} / \textbf{26.42} / \textbf{27.84}&10.96 / 10.96 / 9.32&8.86 / 9.72 / 9.43&16.8 / 17.07 / 16.87 \\
     Average & 23.33 / 22.25 / 20.93&\underline{24.53} / \underline{25.36} / 23.12&12.83 / 12.52 / 11.19&13.9 / 13.54 / 13.16&18.65 / \underline{18.42} / 17.10 \\
        \midrule
    \multicolumn{6}{c}{{\textit{Single-Aspect Prompt}}}\\
    \midrule
     Llama2-7B & -4.02 / -3.32 / -3.9&-1.52 / -2.92 / -5.43&-1.11 / -1.88 / -3.76&0.74 / 2.72 / 4.54&-1.48 / -1.35 / -2.14 \\
    Llama2-13B & 8.39 / 9.5 / 10.91&10.64 / 11.67 / 10.68&-4.44 / -4.52 / -0.96&-0.18 / 1.12 / 0.94&3.60 / 4.44 / 5.39\\
    Qwen1.5-7B & 5.81 / 8.14 / 10.78&5.49 / 5.07 / 6.15&6.26 / 6.35 / 5.07&-1.97 / -1.52 / -2.42&3.9 / 4.51 / 4.89 \\ 
    Qwen1.5-14B & 7.13 / 6.92 / 7.01&22.61 / 22.75 / 22.33& \textbf{28.65} / \textbf{30.71} / \textbf{35.11}&13.88 / 13.68 / 13.94&18.07 / 18.52 / 19.60 \\
    GPT3.5-Turbo & 26.81 / 26.36 / \underline{29.58}&20.62 / 21.22 / 25.01&16.33 / 15.56 / 17.93&9.95 / 7.75 / 6.33&\underline{18.43} / 17.72 / 19.71 \\
    GPT4 & 18.36 / 19.78 / 22.03&20.17 / 21.57 / \underline{23.62}&14.46 / 15.61 / 14.33&7.49 / 5.92 / 3.17&15.12 / 15.72 / 15.79 \\
    \midrule
        \multicolumn{6}{c}{{\textit{Multiple-Aspect Prompt}}}\\
        \midrule
    Llama2-7B & -1.26 / -2.85 / -14.34&-2.2 / -2.59 / -8.87&-3.36 / -7.23 / -16.36&1.74 / 1.99 / 1.82&-1.27 / -2.67 / -9.44 \\
    Llama2-13B & 17.04 / 17.33 / 18.56&4.26 / 3.41 / 10.25&3.59 / 2.1 / 2.24&17.93 / 16.82 / 18.52&10.71 / 9.92 / 12.39 \\
     Qwen1.5-7B & 13.0 / 13.26 / 13.08&11.75 / 11.74 / 15.28&-0.8 / -0.34 / -0.49&10.63 / 9.28 / 15.6&8.65 / 8.49 / 10.87 \\
    Qwen1.5-14B & 25.85 / 26.53 / \textbf{32.28}&18.16 / 18.45 / 22.03&12.25 / 11.32 / 15.26&15.82 / 14.83 / 18.26&18.02 / 17.78 / \underline{21.96} \\
    GPT3.5-Turbo & 26.41 / 26.36 / 28.2&11.16 / 9.86 / 11.38&12.09 / 10.63 / 11.15&\underline{20.93} / \underline{19.56} / \underline{20.78}&17.65 / 16.61 / 17.88 \\
    GPT4 & 
    \textbf{27.26} / \textbf{28.25} / 28.99 &12.68 / 12.17 / 13.26& \underline{20.30} / \underline{22.04} / \underline{24.93} & \textbf{24.05} / \textbf{25.12} / \textbf{27.35}& \textbf{21.07} / \textbf{21.90} / \textbf{23.63} \\

  \bottomrule
\end{tabular}
\end{table*}
\begin{figure*}
  \centering
  \subfigure[GPT4 (M)]{
    \begin{minipage}[b]{1\textwidth}
    \includegraphics[trim=0 2mm 0 2mm, clip,width=\textwidth]{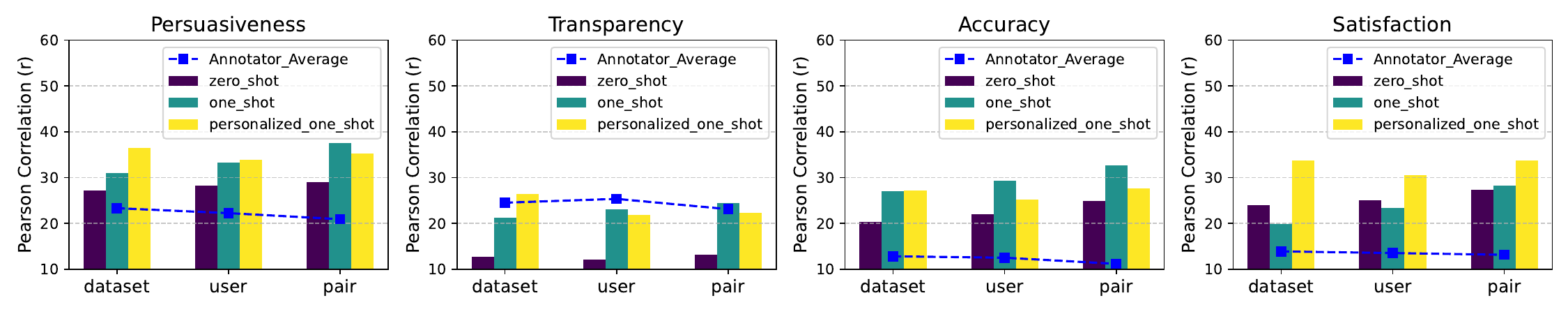}
    \label{fig:shot-GPT4}
    \vspace{-5mm}
    \end{minipage}
}
\vspace{-2mm}
\subfigure[Qwen1.5-14B (M)]{
    \begin{minipage}[b]{1\textwidth}
    \includegraphics[trim=0 2mm 0 3mm, clip,width=1\textwidth]{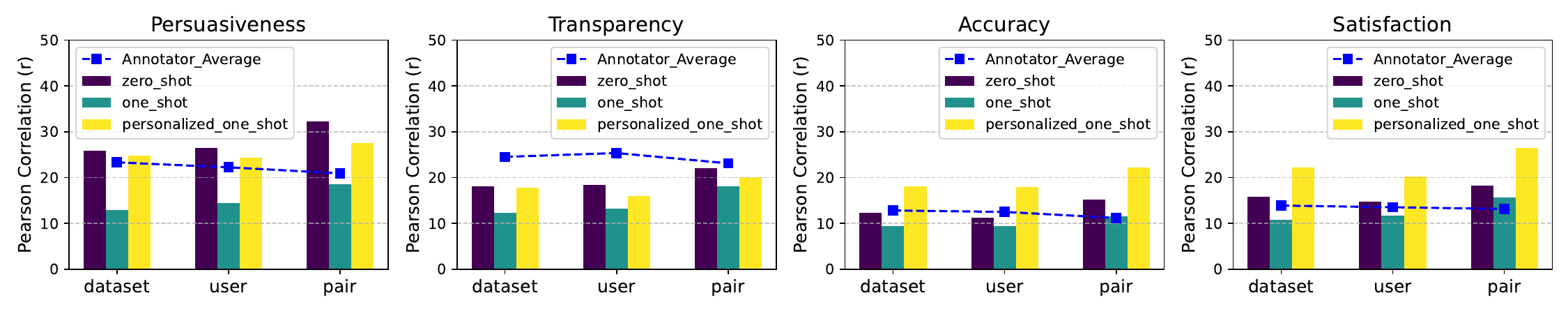}
    \label{fig:shot-Qwen}
    \vspace{-6mm}
    \end{minipage}
}
  \caption{Comparison of 3-level Pearson correlations for zero-shot, (non-personalized) one-shot, and personalized one-shot learning on GPT4(M) and Qwen(M). (M) denotes the multiple-aspect version.}
  \label{fig:one-shot}
  \vspace{-3mm}
\end{figure*}

%% file: 3-RQ1.tex
\section{zero-shot LLM can be a competitive evaluator}

To investigate the quality of assessments given by zero-shot LLMs and answer RQ1, we conduct a 3-level meta-evaluation, with the concrete process introduced in Section \ref{sec:metaevaluation}.
The 5-level scores from real users are used as ground-truth labels.
Zero-shot LLMs instructed by prompts are utilized as evaluators for recommendation explanations.
We calculate Pearson correlations between evaluator results with user labels as the evaluation accuracy.
It should be noted that the evaluation \textbf{accuracy} here refers to the correctness of the assessment given by the evaluator, which is different from the aspect of \textbf{Accuracy} when evaluating recommendation explanation, referring to whether the text is consistent with the user's interests.
We test the accuracy of referenced-based metrics, third-party annotation, and LLM-based evaluator and demonstrate in Table \ref{tab:main}.

\textbf{Experimental Setting.} 
We conduct experiments with 6 LLMs, including Llama2-7B~\cite{llama2}, Llama2-13B~\cite{llama2}, Qwen1.5-7B~\cite{qwen}, Qwen1.5-14B~\cite{qwen}, GPT3.5-Turbo and GPT4~\cite{openai2024gpt4}\footnote{https://openai.com/api/}.
The temperatures of LLMs are set to 0.
The results returned by LLM are parsed to integer scores from 1 to 5.
For the null value in user labels, we set it to 3 since it represents the unknown user attitude.
 For the null value in LLM labels, we set it to 0 since it is usually caused by parsing failures, which should bring penalties to the correlation score.

\textbf{LLMs can achieve evaluation accuracy that is comparable to or surpasses traditional methods.} 
As shown in Table \ref{tab:main}, there are variations in the evaluation accuracy across models. 
GPT-4 demonstrates the highest performance, followed by GPT-3.5 and Qwen1.5-14B, which both show adequate evaluation accuracy. Qwen1.5-7B and Llama2-14B display moderate labeling capabilities, whereas Llama2-7B exhibits poor performance.

In average performance of different aspects, GPT-4 surpasses both third-party annotators and reference-based metrics. GPT3.5 and Qwen1.5 demonstrate comparable accuracy to the third-party annotator, also out-performing reference-based metric.

\textbf{Evaluation accuracies of evaluators are aspect-dependent.}
When analyzing the performance across different aspects, we can see that the third-party annotators show better evaluation accuracy in Persuasiveness and Transparency than Accuracy and Satisfaction.
This implies that Accuracy and Satisfaction are more subject, displaying greater individual variability.
Thus, third-party annotations may not be satisfactory solutions in these aspects.
Experiments indicate that LLMs, particularly GPT-4, perform better in these areas. However, regarding Transparency, LLMs are inferior to human labeling. 
In subsequent sections, we discuss strategies for enhancing the zero-shot LLM evaluator.

\textbf{Results from multiple-aspect prompt v.s. single-aspect prompt.} 
The multiple-aspect involves LLM scoring a text across 4 aspects simultaneously, while in the latter, it assesses each aspect individually.
We find that several LLM evaluators (LLama2-13B, Qwen1.5-7B, GPT3.5, GPT4) perform notably better on Satisfaction when using the multiple-aspect prompt.
This may be due to user satisfaction being a composite consideration of various dimensions.
Thus, scoring on other aspects acts as an implicit Chain-Of-Thought~\cite{NEURIPS2022COT} process that enhances the zero-shot reasoning ability of LLMs~\cite{NEURIPS2022LLMZeroShot}. 
Conversely, the single-aspect prompt yields better results on the Transparency aspect with several LLMs. This may be because the evaluation of Transparency is relatively independent, and separate scoring allows the model to better understand the task without interference from other aspects. 
Overall, we would recommend the multiple-aspect version, as it has no significant gap with single-aspect but requires fewer interactions.
Our experiments on RQ2 and RQ3 are also conducted on multiple-aspect prompts.

\textbf{
Evaluators show varying trends across meta-evaluation levels.} 
The varying trends across levels underscore the importance of selecting the appropriate level based on the objectives of the task.
Human annotators have similar correlations, i.e., evaluation accuracy across three levels.
Reference-based metric primarily demonstrates a trend where results follow Pair-Level > User-Level > Dataset-Level.
A particularly notable example is BLEU-4, which performs even worse than random at the Dataset-Level while performing considerably better at the Pair-Level. 
This discrepancy in results arises since BLEU-4 emphasizes the co-occurrence of 4-gram tokens in target and reference texts. 
Thus, it is influenced by the specific words in the reference text.
At the Dataset-Level, this introduces bias unrelated to user perception.

\label{sec:dataset}
\begin{table}[!]
  \caption{The cost (\$) of evaluating a single text entry on four aspects, where \textbf{(S/M)} denotes single/multiple-aspect prompt.}
  \label{tab:cost}
  \begin{tabular}{cccccc}
    \toprule
    \textbf{Human} & \textbf{GPT3.5 (S)} & \textbf{GPT3.5 (M)} & \textbf{GPT4 (S)} & \textbf{GPT4 (M)}\\
    \midrule
    0.111\$ & 0.00123\$ & 0.000364\$ & 0.0652\$ & 0.0256\$ \\
  \bottomrule
\end{tabular}
\vspace{-5mm}
\end{table}

\textbf{Cost of LLM vs. human annotators.}
We present the expenses associated with evaluating a single text entry across 4 aspects, including both third-party human annotators and closed-source LLMs in Table \ref{tab:cost}.
The utilization of closed-source LLMs (Llama2 and Qwen) incurs no charges.
It can be observed that the cost of multiple-aspect prompt evaluation is lower than that of single-aspect prompt evaluation.
Even for the most costly LLM configuration, i.e., GPT4 (S), the cost remains lower than that of human annotation. 
Thus, we propose that LLM offers a cost-effective solution for evaluating recommendation explanations.

%% file: 4-RQ2.tex

\section{LLM evaluators with in-context learning}
\label{sec:RQ2}
Having investigated zero-shot LLM-based evaluators in Section ~\ref{sec:RQ1}, we would like to further discuss RQ2, i.e., whether LLM can get better results by collaborating with human annotators. 
Concretely, we adopt one-shot learning to exploit human labeling.
The generation of zero-shot and one-shot prompts are detailed in Section \ref{sec:prompt}.
We conduct the experiments on an open-source and a closed-source LLM respectively, which are GPT4 and Qwen1.5-14B. 
Results are shown in Figure \ref{fig:one-shot}, where the one-shot in the legend refers to the non-personalized one-shot.
We can see that whether human labeling can enhance the evaluation accuracy of the LLM is highly dependent on the prompt design and the backbone model used.
Following this, we detail the effects on the two LLMs.

\textbf{Both personalized and non-personalized examples benefit the evaluation accuracy of GPT4.}
As shown in Figure \ref{fig:shot-GPT4}, incorporating human labels helps align the LLM with user preferences, making GPT4(M) perform as well as or better than third-party human annotators across all aspects simultaneously.
Concretely, we find that personalized one-shot learning improves performance over zero-shot on all trails.
Interestingly, non-personalized one-shot learning also yields improvements over zero-shot learning on most trails, though Satisfaction decreases at the Dataset and User Levels.
Therefore, considering collecting real user labeling on public datasets is impractical, our experiments demonstrate that for GPT-4, labels from other users can still guide LLMs in evaluating user perceptions.

\textbf{Personalized cases facilitate GPT-4 in learning user scoring bias.}
When comparing the results of GPT4 across different learning strategies and three levels, we notice that for zero-shot and non-personalized one-shot, the Pair-Level evaluation accuracy consistently surpasses that of the Dataset-Level, but this is not the case for personalized one-shot learning.
This indicates that while the evaluator based on zero-shot and non-personalized one-shot learning can distinguish between recommendation texts generated for the same user-movie pair, it may not be good at capturing biases inherent in user rating, such as higher overall ratings from one user compared to another. Introducing personalized examples can mitigate the issue.

\textbf{Non-personalized one-shot learning does not work well on Qwen1.5-14B.}
We observe that non-personalized one-shot learning does not effectively improve and even impairs the performance of Qwen1.5-14B as the evaluator, as shown in Figure \ref{fig:shot-Qwen}. This may be because single-shot prompts from other users introduce additional bias into Qwen1.5-14B's ratings. Personalized one-shot learning brings improvements in Accuracy and Satisfaction. These two aspects, as mentioned in Section \ref{sec:RQ1}, exhibit greater individual variability than others. Introducing personalized information helps Qwen1.5-14B better capture subjective user perception.

In summary, personalized one-shot learning can effectively enhance the evaluation accuracy of GPT4 and on Accuracy and Satisfaction of Qwen1.5-14B. 
Nevertheless, the process of collecting corresponding user labels is often challenging, particularly on publicly available datasets.
As an alternative solution, the incorporation of labels from other users can also enhance the evaluation provided by GPT-4. 
Despite this, human annotation remains relatively expensive. 
In the next section, we discuss how to improve the accuracy and stability of evaluations without human labeling.

%% file: 5-RQ3.tex
\begin{figure*}[t]
    \centering
    \includegraphics[width=1\linewidth]{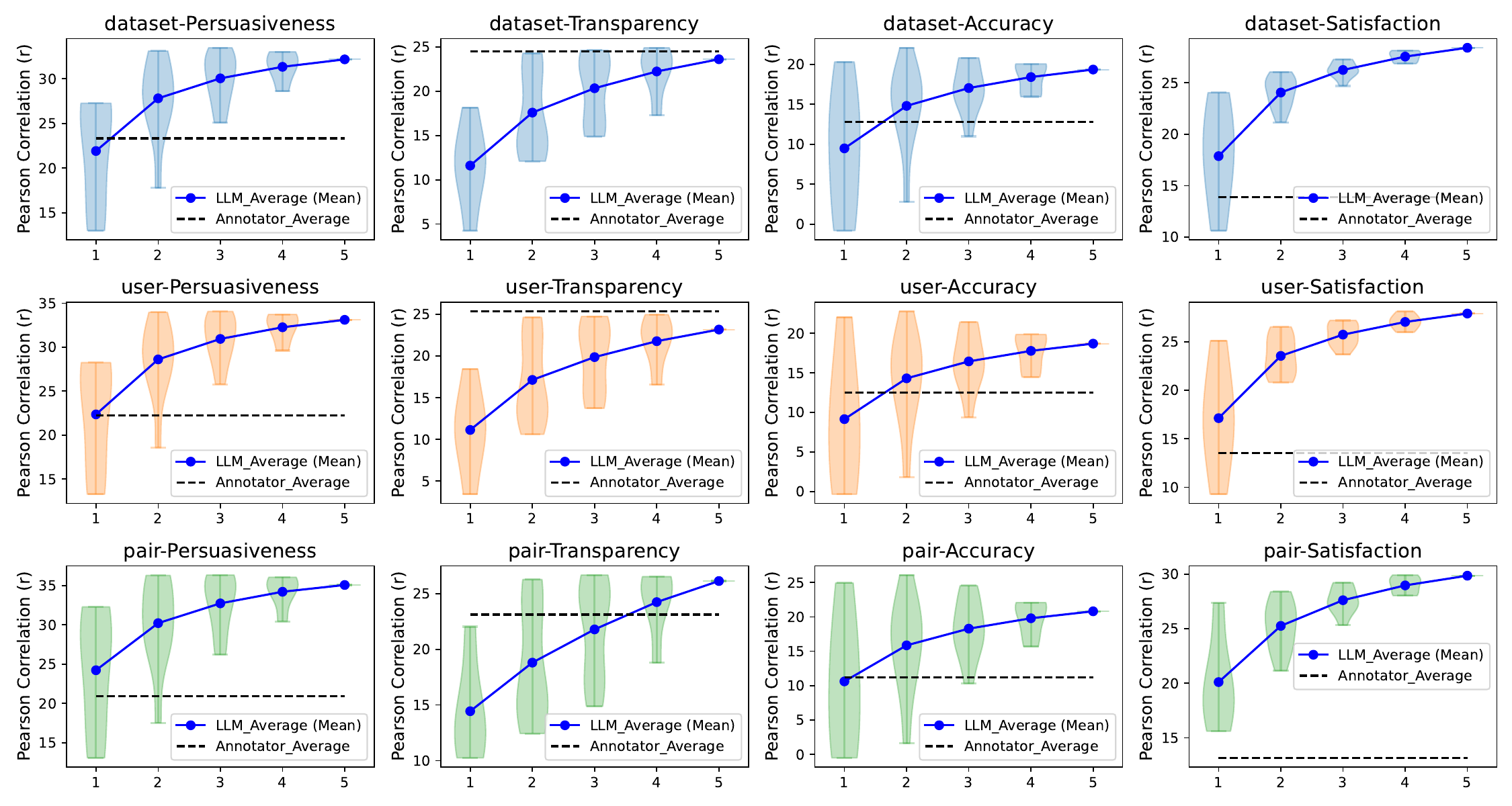}
    \vspace{-5mm}
    \caption{Distribution of evaluation accuracy from ensemble results of \#N LLM evaluators. Values on the x-axis denote \#N. }
    \label{fig:ensemble}
    
\end{figure*}
\section{Two heads are better than one}
\label{sec:RQ3}
In RQ1, although LLMs can achieve comparable evaluation accuracy on average of tested aspects, we find that the evaluation accuracy of LLMs varies by aspect.
For instance, the evaluation accuracy of GPT-4 on Transparency is less ideal compared to other aspects.
This raises the question of how to apply LLM evaluators to untested data or aspects to ensure the evaluation quality.
Inspired by the common method utilized in human annotation to average the ratings given by multiple annotators, we ensemble results from various LLMs.
We explore RQ3, conducting experiments on 5 LLM evaluators from RQ1 (excluding Llama2-7B).
We ensemble multiple LLMs by averaging their ratings to obtain the final scores.
In Figure \ref{fig:ensemble}, the x-axis represents the number of LLMs included in the ensemble and the y-axis indicates the corresponding evaluation accuracy.

\textbf{Ensemble of multiple LLMs improves evaluation accuracy and stability.}
In Figure \ref{fig:ensemble}, we can see that the expectation of evaluation accuracy (mean) increases with \#N on all level aspects. 
In addition, the lower bound of the evaluation accuracy also rises.
This suggests that ensemble multiple LLMs can mitigate the issue of a single evaluator performing poorly on certain aspects, such as Qwen1.5-7B on Accuracy aspect.

\textbf{The upper bound of evaluation accuracy  decreases as \#N rises on Accuracy aspect.}
We notice that while the expectation (mean) of accuracy increases with \#N, the upper bound of evaluation accuracy starts to decline when \#N $\geq 3$ on the Accuracy aspect.
This may be due to the subjective nature of Accuracy aspect, which results in suboptimal outcomes when there are too many evaluators.
Another possible reason is that, as observed in Table \ref{tab:main}, two LLMs (Llama2-13B and Qwen1.5-7B) perform relatively poor on the Accuracy aspect, which could negatively impact the effectiveness of LLM ensembles.

In summary, when dealing with untested datasets and aspects, we recommend aggregating zero-shot LLM evaluators to ensure more stable evaluations.

%% file: 6-RelatedWork.tex
\section{Related Work}
\textbf{Evaluating Explainable Recommendation}
Evaluation of explainable recommendations has been an important topic.
In previous studies, widely used evaluation approaches include online user study, third-party annotation, and reference-based metrics.
An online user study is the most accurate method to evaluate user perceptions. In Ex3~\cite{Ex3}, authors deploy their model online and observe an increase in traffic.
User studies are also utilized to help gain insights about explainable recommendations. ~\citet{UserPerceptionTois} track users' intentions, expectations, and experiences in interactions with an explainable recommendation system.
~\citet{ConflictGoalsOfExplanations} investigate the relationship between various goals in explainable recommendations.
The limitation of this line of evaluation is that it is hard to acquire, especially on public datasets.
Researchers resort to third-party annotations as human labels.
Some utilize crowdsourcing to collect labels~\cite{ExpScore2022WWW} and others employ experienced annotators~\cite{chen2021eCr,lei2023recexplainer}.
Although they are easier to obtain compared to real user labels, human labels are still expensive.
In addition, our experimental results find that third-party annotations may be less accurate on aspects that are highly subjective.
Reference-based metrics, e.g. BLEU~\cite{DBLP:bleu2022ACL}, ROUGE~\cite{Rouge2004Lin} and their variants~\cite{sellam2020bleurt} are utilized by calculating the similarity of target texts with reference texts.
BLEU and ROUGE are almost the most common methods for evaluating text-based recommendation explanations~\cite{li2022personalized,li2021personalizedTransformer}.
These quantitative metrics are easy to acquire and have a standard calculation process.
This line of methods is limited to datasets with reference text attached, i.e., self-explanations or reviews.

Studies exist proposing and utilizing novel evaluation methods for explainable recommendations, each with its own advantages and limitations.
~\citet{agnostic} propose a model agnostic framework that evaluates explanations from the aspects of faithfulness and scrutability.
ExpScore~\cite{ExpScore2022WWW} design models to generate evaluations from human labels. 
These methods have limited applicability or require collecting a certain quantity of human labels.
RecExplainer~\cite{lei2023recexplainer} utilizes both human and LLM as annotators for generated explanations.
To the best of our knowledge, we are the first to conduct a meta-evaluation on LLM for recommendation explanations to comprehensively study its capability for the task.

\textbf{LLM-based NLG Evaluation}
The emergence of LLM has sparked interest in its potential applications in evaluating NLG tasks.
~\citet{ChatGPTourperformsCrowdWorker} investigate that ChatGPT outperforms crowd workers on annotating various NLG tasks.
Previous studies find that LLM can be effective annotators on various tasks when prompting appropriately~\cite{fu2023gptscore,kocmi2023LLMEvaluateTransaltion,wang2023chatgpt,wang2023automatedpersonaliztion}.
In-context learning (ICL) is also employed to generate annotations using few-shot learning~\cite{brown2020languagearefewshotlearner,shin2021few-shotSemantic,rubin2022learning}.
Researchers summarize the advancements in the field into surveys~\cite{LLMSurvey1,LLMSurvey2}.
Among these studies, the one most similar to ours is the evaluation of personalized text generation, e.g., reviews, comments on social media, etc~\cite{wang2023automatedpersonaliztion}.
Our work adds to these studies by utilizing LLMs as evaluators for recommendation explanation texts.

%% file: 7-Conclusion.tex
\section{Conclusion}
In this paper, we investigate the feasibility of utilizing LLMs as evaluators for recommendation explanation texts.
We leverage real user feedback as ground-truth labels to validate the quality of LLM evaluation.
Our studies consider zero-shot LLM evaluation, collaborating with human labels and the ensemble of multiple LLMs. 
Our key findings include 1) some LLMs, such as GPT-4, can achieve evaluation accuracy comparable to or better than traditional methods; 2) GPT4 can effectively learn preference from human labels; 3) when applying to untested datasets and aspects, aggregating multiple heterogeneous zero-shot LLMs is recommended to improve the accuracy and stability of the evaluation.
We propose that LLM can be a reproducible and cost-effective solution for evaluating recommendation explanations.
As a preliminary investigation into the meta-evaluation of LLM on recommendation explanations, our work is limited to text-based explanations. In the future, unified evaluation protocols that encompass a broader range of explanation formats can be studied. In addition, developing novel methodologies to further enhance the evaluation accuracy of LLMs is also an important area worth considering.